\documentclass[prl,aps,twocolumn,superscriptaddress,showpacs,preprintnumbers,
               amsmath,amssymb,floatfix]{revtex4}

\def\gtwid{\mathrel{\raise.3ex\hbox{$>$\kern-.75em\lower1ex\hbox{$\sim$}}}}
\def\ltwid{\mathrel{\raise.3ex\hbox{$<$\kern-.75em\lower1ex\hbox{$\sim$}}}}

\usepackage{graphicx} 
\usepackage{dcolumn}  
\usepackage{bm}       

\begin{document}

\title{Analysis of the low-energy electron-recoil spectrum of the CDMS experiment}

\affiliation{Department of Physics, California Institute of Technology, Pasadena, CA 91125, USA}
\affiliation{Department of Physics, Case Western Reserve University, Cleveland, OH  44106, USA}
\affiliation{Fermi National Accelerator Laboratory, Batavia, IL 60510, USA}
\affiliation{Lawrence Berkeley National Laboratory, Berkeley, CA 94720, USA}
\affiliation{Department of Physics, Massachusetts Institute of Technology, Cambridge, MA 02139, USA}
\affiliation{Department of Physics, Queen's University, Kingston, ON, Canada, K7L 3N6}
\affiliation{Department of Physics, Saint Olaf College, Northfield, MN  55057}
\affiliation{Department of Physics, Santa Clara University, Santa Clara, CA 95053, USA}
\affiliation{Department of Physics, Stanford University, Stanford, CA 94305, USA}
\affiliation{Department of Physics, Syracuse University, Syracuse, NY 13244}
\affiliation{Department of Physics, Texas A\&M University, College Station, TX 93106, USA} 
\affiliation{Department of Physics, University of California, Berkeley, CA 94720, USA}
\affiliation{Department of Physics, University of California, Santa Barbara, CA 93106, USA}
\affiliation{Departments of Phys. \& Elec. Engr., University of Colorado Denver, Denver, CO 80217, USA}
\affiliation{Department of Physics, University of Florida, Gainesville, FL 32611, USA}
\affiliation{School of Physics \& Astronomy, University of Minnesota, Minneapolis, MN 55455, USA}
\affiliation{Physics Institute, University of Z\"urich, Z\"urich, Switzerland}

\author{Z.~Ahmed} \affiliation{Department of Physics, California Institute of Technology, Pasadena, CA 91125, USA}
\author{D.S.~Akerib} \affiliation{Department of Physics, Case Western Reserve University, Cleveland, OH  44106, USA} 
\author{S.~Arrenberg} \affiliation{Physics Institute, University of Z\"urich, Z\"urich, Switzerland} 
\author{C.N.~Bailey} \affiliation{Department of Physics, Case Western Reserve University, Cleveland, OH  44106, USA} 
\author{D.~Balakishiyeva}\affiliation{Department of Physics, University of Florida, Gainesville, FL 32611, USA}
\author{L.~Baudis} \affiliation{Physics Institute, University of Z\"urich, Z\"urich, Switzerland} 
\author{D.A.~Bauer} \affiliation{Fermi National Accelerator Laboratory, Batavia, IL 60510, USA} 
\author{J.~Beaty} \affiliation{School of Physics \& Astronomy, University of Minnesota, Minneapolis, MN 55455, USA} 
\author{P.L.~Brink} \affiliation{Department of Physics, Stanford University, Stanford, CA 94305, USA} 
\author{T.~Bruch} \affiliation{Physics Institute, University of Z\"urich, Z\"urich, Switzerland} 
\author{R.~Bunker} \affiliation{Department of Physics, University of California, Santa Barbara, CA 93106, USA} 
\author{B.~Cabrera} \affiliation{Department of Physics, Stanford University, Stanford, CA 94305, USA} 
\author{D.O.~Caldwell} \affiliation{Department of Physics, University of California, Santa Barbara, CA 93106, USA} 
\author{J.~Cooley} \affiliation{Department of Physics, Stanford University, Stanford, CA 94305, USA} 
\author{P.~Cushman} \affiliation{School of Physics \& Astronomy, University of Minnesota, Minneapolis, MN 55455, USA} 
\author{F.~DeJongh} \affiliation{Fermi National Accelerator Laboratory, Batavia, IL 60510, USA} 
\author{M.R.~Dragowsky} \affiliation{Department of Physics, Case Western Reserve University, Cleveland, OH  44106, USA} 
\author{L.~Duong} \affiliation{School of Physics \& Astronomy, University of Minnesota, Minneapolis, MN 55455, USA} 
\author{E.~Figueroa-Feliciano}\affiliation{Department of Physics, Massachusetts Institute of Technology, Cambridge, MA 02139, USA}
\author{J.~Filippini} \affiliation{Department of Physics, University of California, Berkeley, CA 94720, USA}\affiliation{Department of Physics, California Institute of Technology, Pasadena, CA 91125, USA} 
\author{M.~Fritts} \affiliation{School of Physics \& Astronomy, University of Minnesota, Minneapolis, MN 55455, USA} 
\author{S.R.~Golwala} \affiliation{Department of Physics, California Institute of Technology, Pasadena, CA 91125, USA} 
\author{D.R.~Grant} \affiliation{Department of Physics, Case Western Reserve University, Cleveland, OH  44106, USA} 
\author{J.~Hall} \affiliation{Fermi National Accelerator Laboratory, Batavia, IL 60510, USA} 
\author{R.~Hennings-Yeomans} \affiliation{Department of Physics, Case Western Reserve University, Cleveland, OH  44106, USA} 
\author{S.~Hertel}\affiliation{Department of Physics, Massachusetts Institute of Technology, Cambridge, MA 02139, USA}
\author{D.~Holmgren} \affiliation{Fermi National Accelerator Laboratory, Batavia, IL 60510, USA} 
\author{L.~Hsu} \affiliation{Fermi National Accelerator Laboratory, Batavia, IL 60510, USA} 
\author{M.E.~Huber} \affiliation{Departments of Phys. \& Elec. Engr., University of Colorado Denver, Denver, CO 80217, USA} 
\author{O.~Kamaev} \affiliation{School of Physics \& Astronomy, University of Minnesota, Minneapolis, MN 55455, USA} 
\author{M.~Kiveni} \affiliation{Department of Physics, Syracuse University, Syracuse, NY 13244} 
\author{M.~Kos} \affiliation{Department of Physics, Syracuse University, Syracuse, NY 13244} 
\author{S.W.~Leman} \affiliation{Department of Physics, Massachusetts Institute of Technology, Cambridge, MA 02139, USA} 
\author{R.~Mahapatra}\affiliation{Department of Physics, Texas A\&M University, College Station, TX 93106, USA} 
\author{V.~Mandic} \affiliation{School of Physics \& Astronomy, University of Minnesota, Minneapolis, MN 55455, USA} 
\author{D.~Moore} \affiliation{Department of Physics, California Institute of Technology, Pasadena, CA 91125, USA}
\author{K.A.~McCarthy}\affiliation{Department of Physics, Massachusetts Institute of Technology, Cambridge, MA 02139, USA}
\author{N.~Mirabolfathi} \affiliation{Department of Physics, University of California, Berkeley, CA 94720, USA} 
\author{H.~Nelson} \affiliation{Department of Physics, University of California, Santa Barbara, CA 93106, USA} 
\author{R.W.~Ogburn} \affiliation{Department of Physics, Stanford University, Stanford, CA 94305, USA}\affiliation{Department of Physics, California Institute of Technology, Pasadena, CA 91125, USA} 
\author{M.~Pyle} \affiliation{Department of Physics, Stanford University, Stanford, CA 94305, USA}
\author{X.~Qiu} \affiliation{School of Physics \& Astronomy, University of Minnesota, Minneapolis, MN 55455, USA} 
\author{E.~Ramberg} \affiliation{Fermi National Accelerator Laboratory, Batavia, IL 60510, USA} 
\author{W.~Rau} \affiliation{Department of Physics, Queen's University, Kingston, ON, Canada, K7L 3N6}
\author{A.~Reisetter} \affiliation{Department of Physics, Saint Olaf College, Northfield, MN  55057}\affiliation{School of Physics \& Astronomy, University of Minnesota, Minneapolis, MN 55455, USA}
\author{T.~Saab}\affiliation{Department of Physics, University of Florida, Gainesville, FL 32611, USA}
\author{B.~Sadoulet} \affiliation{Lawrence Berkeley National Laboratory, Berkeley, CA 94720, USA}\affiliation{Department of Physics, University of California, Berkeley, CA 94720, USA}
\author{J.~Sander} \affiliation{Department of Physics, University of California, Santa Barbara, CA 93106, USA} 
\author{R.W.~Schnee} \affiliation{Department of Physics, Syracuse University, Syracuse, NY 13244} 
\author{D.N.~Seitz} \affiliation{Department of Physics, University of California, Berkeley, CA 94720, USA} 
\author{B.~Serfass} \affiliation{Department of Physics, University of California, Berkeley, CA 94720, USA} 
\author{K.M.~Sundqvist} \affiliation{Department of Physics, University of California, Berkeley, CA 94720, USA} 
\author{G.~Wang} \affiliation{Department of Physics, California Institute of Technology, Pasadena, CA 91125, USA}
\author{S.~Yellin} \affiliation{Department of Physics, Stanford University, Stanford, CA 94305, USA} \affiliation{Department of Physics, University of California, Santa Barbara, CA 93106, USA}
\author{J.~Yoo} \affiliation{Fermi National Accelerator Laboratory, Batavia, IL 60510, USA} 
\author{B.A.~Young} \affiliation{Department of Physics, Santa Clara University, Santa Clara, CA 95053, USA}

\collaboration{CDMS Collaboration}
\noaffiliation

\begin{abstract}
We report on the analysis of the low-energy electron-recoil spectrum from the CDMS~II experiment using data with an exposure of 443.2 kg-days. The analysis provides details on the observed counting rate and possible background sources in the energy range of 2\,-\,8.5 keV. We find no significant excess in the counting rate above background, and compare this observation to the recent DAMA results. In the framework of a conversion of a dark matter particle into electromagnetic energy, our 90\% confidence level upper limit of \mbox{$0.246$ events/kg/day} at 3.15\,keV is lower than the total rate above background observed by DAMA by 8.9$\sigma$. In absence of any specific particle physics model to provide the scaling in cross section between NaI and Ge, we assume a Z$^2$ scaling. With this assumption the observed rate in DAMA differs from the upper limit in CDMS by 6.8$\sigma$. Under the conservative assumption that the modulation amplitude is 6\% of the total rate we obtain upper limits on the modulation amplitude a factor of $\sim$2 less than observed by DAMA, constraining some possible interpretations of this modulation.
\end{abstract}
\preprint{FERMILAB-PUB-09-340-E}
\maketitle

Astrophysical observations strongly suggest that non-luminous, non-baryonic matter constitutes most of the matter in the Universe. This dark matter should be locally distributed in dark halos of galaxies such as the Milky Way, enabling the direct detection of the dark matter particles via their interaction in terrestrial detectors. The movement of the Earth around the Sun would provide an annual modulation of the counting rate, caused by the change in the relative velocity of the dark matter particles and the earthbound target. The DAMA collaboration claims the observation of such a modulation, in two different NaI(Tl) scintillation detector arrays, the original DAMA/NaI setup \cite{dama2000} and the upgraded DAMA/LIBRA experiment \cite{damalibra2008}. The observed signal is in the 2\,-\,6\,keV electron-equivalent energy range with a periodicity of 0.998$\pm$0.003 years and a phase of 144$\pm$8 days. The DAMA collaboration claims that no known systematic detector effect could explain the modulation signal. The modulation phase is consistent with the expected signature of galactic dark matter particles interacting in a terrestrial detector. However, the original interpretation of the DAMA result as a signal from Weakly Interacting Massive Particles (WIMPs) that would interact via nuclear recoils \cite{dama2000} is inconsistent with other experimental results \cite{cdms2008, xenon10,xenon10sd,coupp,kims} \footnote{There are scenarios, such as micro-channeling effects of ions in crystals, which still leave a little room at low WIMP masses ($\sim$10\,GeV range)~\cite{channel2007,damachannel2007}.}. Note, that the DAMA detectors do not discriminate between electron recoils and nuclear recoils. 

A signal from an electromagnetic dark matter interaction should be detectable in the Cryogenic Dark Matter Search (CDMS) experiment, but would be rejected in our standard search for nuclear recoils \cite{cdms2008}. The possibility of an electron-recoil signal from axion-like dark matter particles has recently been investigated \cite{damaaxion,cogent,cdmsaxion2008}. In this paper, we present a general analysis of our low-energy electron-recoil spectrum, provide details on the observed counting rate in this energy range, and comment on the implications of these results on possible interpretations of the energy spectrum and the modulation signal observed by DAMA.

The CDMS collaboration operates a total of 19 Ge and 11 Si crystal detectors, each having a mass of $\sim$250\,g and $\sim$100\,g, respectively, at a temperature of $\sim$\,40\,mK in the Soudan Underground Laboratory~\cite{zips, prd118}. The ionization and phonon energy of every event is read out simultaneously. The recoil energy is reconstructed from them. The ratio of ionization to recoil energy, the ionization yield, discriminates nuclear- from electron-recoil events.

In this analysis we consider data with a total exposure of 443.2 kg-days before cuts, which has been acquired in two run periods between October 2006 and July 2007 (designated as R123 and R124) and is the same dataset used for an axion search analysis \cite{cdmsaxion2008}. Three of the 19 Ge detectors were excluded because of readout failures and another one due to reduced trigger performance at low energies.  From the remaining 15 Ge detectors one suffered from reduced trigger performance in R123 and two from incomplete neutralization in R124 which have also been left out of the analysis. The silicon detectors were not considered. We required that an event had to pass several cuts. The events needed to have an ionization energy at least 3$\sigma$ above the mean noise and be recorded in only one detector. All 30 detectors were used to select these single-scatter events. Moreover, we demanded that there was no signal in the scintillator veto shield surrounding the detectors. The length of the veto coincidence window was set to 50 $\mu$s. In order to explore the low-energy electron-recoil spectrum we selected events inside the  2$\sigma$ electron-recoil band in ionization yield \cite{cdms2008}. The fiducial volume was measured using nuclear-recoil events from calibrations with a $^{252}$Cf source because of the uniform distribution of neutrons throughout the detector. We excluded all datasets taken within 3 days after a neutron calibration to avoid high gamma rates due to neutron activation of the detectors' supporting structure made predominantly of copper. The remnant rate of $^{64}$Cu contributes less than 2\% to the mean counting rate at low energies and decreases with a half life of 12.7\,h.

\begin{figure}[t!]
\includegraphics[width=3.4in]{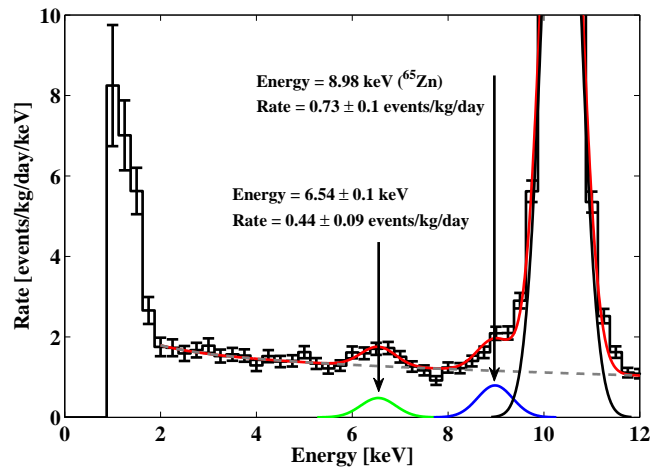}
\caption{\small Fit (red line) to the efficiency corrected low-energy spectrum consisting of a background model (gray/dashed) and three Gaussian distribution functions describing the 10.36\,keV line from $^{71}$Ge (black), the 8.98\,keV line from $^{65}$Zn (blue) and a line at the energy of $^{55}$Mn (green, see text). The total counting rate of the latter two lines is given in the figure.}
\label{fig:enespc}
\end{figure}

The summed background spectrum of all considered detectors, taking into account the detection efficiency \cite{cdmsaxion2008}, is shown in Fig.\,\ref{fig:enespc}. For reference, the \mbox{corresponding} counting rates are also given in Table\,\ref{tab:lowenergycounts}. In this analysis we consider the electron-equivalent energy range between 2 and 8.5\,keV based on the ionization signal, in which the background rate is \mbox{$\sim$1.5 events/kg/day/keV}. Fig.\,\ref{fig:enespc} also illustrates a simple fit to the observed electron-recoil spectrum. The fit incorporates known spectral lines at 10.36\,keV and 8.98\,keV, both outside of our analysis window. The former is caused by X-rays and Auger-electrons from the decay of $^{71}$Ge, a product of neutron capture on $^{70}$Ge during neutron calibrations. The latter originates in the decay of remnant $^{65}$Zn from cosmogenic activation of the detectors. We also fit for a spectral line corresponding to an excess of events observed near 6.5\,keV, which is likely caused by the de-excitation of $^{55}$Mn; this feature is discussed further below. Each peak is fit by a Gaussian distribution function with width fixed at CDMS's measured energy resolution \cite{cdmsaxion2008}. The detector-averaged r.m.s. energy resolution $\sigma(E)$ below 10 keV is given by:
\begin{equation}
 \sigma(E)=\sqrt{(0.293)^2+(0.056)^2 \, E} \,  \textrm{[keV]}
\end{equation}
\noindent
where $E$ is the measured energy in keV.

\begin{table}[b!]
\begin{tabular}{|c|c||c|c||c|c|}
\hline
 Energy & Rate & Energy & Rate &Energy & Rate\\
\hline
2.0& 1.93$\pm$0.24& 4.25&1.52$\pm$0.15& 6.5&1.70$\pm$0.15\\
2.25& 1.96$\pm$0.22&4.5&1.50$\pm$0.15&  6.75&1.84$\pm$0.16 \\
2.5& 1.63$\pm$0.19&4.75&1.55$\pm$0.15&  7.0&1.43$\pm$0.14   \\
2.75&1.73$\pm$0.18&5.0&1.52$\pm$0.15&    7.25&1.47$\pm$0.14 \\
3.0&2.04$\pm$0.19&5.25&1.43$\pm$0.14&     7.5&1.26$\pm$0.13 \\
3.25&1.40$\pm$0.15&  5.5&1.32$\pm$0.13 &  7.75&1.03$\pm$0.12  \\
3.5&1.70$\pm$0.17&  5.75&1.19$\pm$0.13&  8.0&1.29$\pm$0.13  \\
3.75&1.65$\pm$0.16& 6.0&1.75$\pm$0.15&  8.25&1.31$\pm$0.13  \\
4.0&1.41$\pm$0.15&  6.25&1.73$\pm$0.15&  8.5&1.40$\pm$0.13   \\
\hline
\end{tabular}
\caption{\small Rate [events/kg/day/keV] in the 2 - 8.5 keV energy range. }
    \label{tab:lowenergycounts}
\end{table}

$^{55}$Mn can be produced by electron capture of remnant $^{55}$Fe from cosmogenic activation. The de-excitation of $^{55}$Mn results in a spectral line at 6.54\,keV, matching exactly the energy of the corresponding peak in our spectrum. While at the surface the detectors were exposed to fast neutrons from cosmic-ray showers. Gamma-rays from isotopes produced in Ge by these fast cosmic-ray neutrons have been observed in the CoGeNT experiment, which uses a p-type contact germanium detector providing an excellent energy resolution \cite{cogent}. The most dominant lines in their spectrum are from $^{65}$Zn with an energy of 8.98 keV and $^{68,71}$Ge with an energy of 10.36 keV, which are both also visible in our spectrum. Calculations of the production rate of cosmogenic isotopes show that $^{55}$Fe is produced in Ge \cite{mei2009}. The long halflife of  $^{55}$Fe of 2.73\,y allows a remaining activity of this isotope in the detectors. Since the activation stopped when the detectors were moved underground, the time evolution of this counting rate would enable us to determine if it is caused by $^{55}$Fe isotopes. However, the uncertainties in the production rate and on the time the detectors spent at the surface are likely too great to give a reliable constraint on the total rate expected from the de-excitation of $^{55}$Mn.

We carried out a profile likelihood analysis in order to search for an excess of event rate above background \cite{james}. The event rate per unit measured energy ($E$) and per detector ($d$) including background was written as:
\begin{eqnarray}
  R(E,d) = B(E,d) + A(E,d)
\label{eqn:rate}
\end{eqnarray}
\noindent 
The background $B(E,d)$ is assumed to be of the form 
\begin{eqnarray}
 \nonumber
B(E,d)&=&\varepsilon(E,d) \cdot \left[C(d)+D(d)E+\frac{H(d)}{E} \right] \\
     &+&\eta \cdot \varepsilon(E,d) \cdot \frac{\lambda_{6.54}}{\sqrt{2 \pi} \sigma_{6.54}(d)} e^{-\left(\frac{E-6.54}{\sqrt{2} \sigma_{6.54}(d)}\right)^2} 
\label{background}
\end{eqnarray}
\noindent
where $C(d)$, $D(d)$ and $H(d)$ are free parameters determined by the fit routine and $\varepsilon(E,d)$ is the energy-dependent detection efficiency. The Gaussian represents a contribution from $^{55}$Fe decays at an energy of 6.54\,keV. $A(E,d)$ represents a spectral line at a given energy $E_0$.
Thus, we used a Gaussian distribution function multiplied with the efficiency:
\begin{eqnarray}
A(E,d) = \varepsilon(E,d) \cdot \frac{\lambda_{0}}{\sqrt{2 \pi} \sigma_{0}(d)} e^{-\left(\frac{E-E_0}{\sqrt{2} \sigma_{0}(d)}\right)^2}
\label{specline}
\end{eqnarray}
\noindent
Since we have no constraint on the $^{55}$Fe contribution to the spectrum we do not subtract a possible background contribution. The reason for introducing the additional factor $\eta$ in \eqref{background} is that, while scanning over the recoil energy and approaching the 6.54\,keV background peak, the fit function actually consists of a sum of two Gaussians at the same energy. Thus, it serves as a weight  suppressing the importance of the $^{55}$Fe rate in the background model $B(E,d)$. We varied $\eta$ in steps of 0.1 between 0 and 1 and took the most conservative of these limits for each energy.

\begin{figure}[t!]
\includegraphics[width=3.4in]{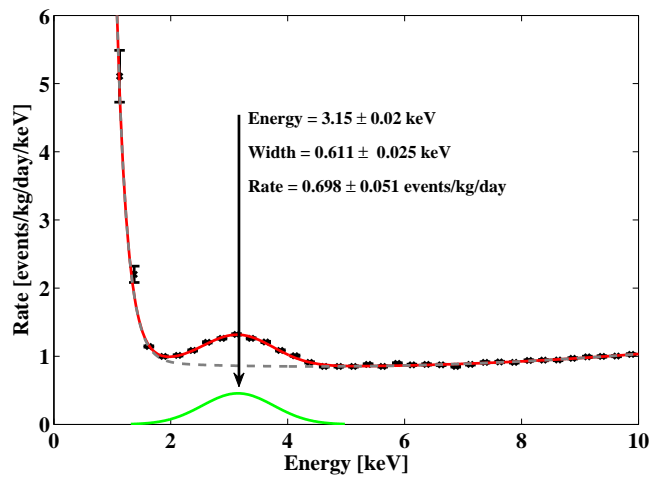}
\caption{\small Fit (red) to the published DAMA/LIBRA low-energy spectrum \cite{damalibra2008}, consisting of a background model (grey/dashed) and a Gaussian distribution function (green). The parameters of the Gaussian are given in the figure.}
\label{fig:damaspectrumfit}
\end{figure}

The fit was performed by a maximization of the unbinned log-likelihood function 
\begin{eqnarray}
  \log(\mathcal{L}) = - R_T + \sum_{i,j} \log R(E_i,d_j)
\label{eqn:loglike}
\end{eqnarray}
\noindent
where the sum goes over events ($i$) and detectors ($j$), with respect to $\lambda_{0}$ and the background parameters. $R_T$ denotes the total sum of the event rate ($R$) over energy and all detectors. We find no statistically significant excess of the event rate above background.  We set a Bayesian 90\% confidence level (CL) upper limit on the total counting rate $\lambda_0$ by integrating the profile likelihood function in the physically allowed region ($\lambda_0>0$). 

\begin{figure}[t!]
\includegraphics[width=3.4in]{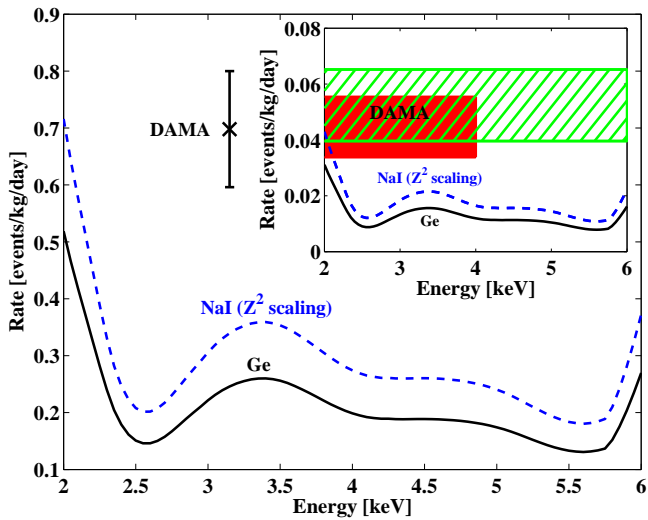}
\caption{\small 90\% CL upper limit on the total counting rate in Ge from this analysis (black/solid). The corresponding upper limit on the total counting rate in NaI under the assumption of a Z$^2$ scaling of the conversion cross section (see text) is also shown (blue/dashed). The black data point with \mbox{2$\sigma$} error bars gives the total counting rate of the 3.15\,keV peak of DAMA/LIBRA derived from a fit to their spectrum (see Fig.\,\ref{fig:damaspectrumfit}). The insert compares the upper limit on the modulation amplitude assumed to be 6\% of the unscaled upper limit (black/solid) and the Z$^2$ scaled upper limit in NaI (blue/dashed) with the 2$\sigma$ regions of the annual modulation amplitude observed by DAMA (NaI+LIBRA) in the 2\,-\,4 keV (red/filled) and 2\,-\,6 keV (green/hatched) energy range.}
\label{fig:upperlim}
\end{figure}

The annual modulation signature observed by DAMA \cite{damalibra2008} may be interpreted as the conversion of a dark matter particle into electromagnetic energy in the detector. In this case it should be possible to observe the corresponding signal in the electron-recoil spectrum of CDMS. The upper limits on an excess rate presented in this paper should thus help to identify or constrain possible models which can explain the annual modulation signature observed by DAMA. The total counting rate above background observed by DAMA/LIBRA in the claimed signal region has been obtained from a fit to their spectrum consisting of a Gaussian and a background model shown in Fig. \ref{fig:damaspectrumfit} giving a total rate of \mbox{$0.698\pm0.051$ events/kg/day}. A direct comparison between the 90\% CL upper limits from this analysis (black/solid) and the total rate above background observed by DAMA (black data point with 2$\sigma$ error bars in the figure) is shown in \mbox{Fig.\,\ref{fig:upperlim}}. At the energy of the DAMA peak (3.15\,keV) the observed rate is 8.9$\sigma$ away from the upper limit on the rate in CDMS of $0.246$ events/kg/day. Though the peak of \mbox{Fig.\,\ref{fig:damaspectrumfit}} may contain a contribution from the decay of $^{40}$K and the subsequent de-excitation of $^{40}$Ar resulting in a spectral line at 3.2\,keV, no information is supplied on the actual rate of such a background \cite{damaapparatus}. Thus, no subtraction is performed, which would reduce the difference between the upper limit from CDMS and the excess rate in DAMA.

The event rates in the CDMS and DAMA detection media may differ depending on the coupling of the dark matter particle. Thus, the upper limits in Ge have to be scaled to the expected rate in NaI in order to perform a comparison in a particular model. For an electromagnetic conversion of a dark matter particle, the particle velocity is essentially irrelevant (in contrast to the calculation for nuclear recoils, where the energy threshold provides a minimum velocity for the phase space integral). Thus, the annual modulation signature is only caused by a change in the particle flux over the course of the year. The total counting rates per unit mass of such a conversion in the case of a Ge and a NaI target are related by the following condition:
\begin{equation}
\frac{R_{NaI}}{R_{Ge}}= \frac{A_{Ge}}{A_I + A_{Na}} \cdot \frac{\sigma_{I} + \sigma_{Na}}{ \sigma_{Ge}}
\end{equation}
\noindent
where $A_{i}$ is the atomic mass of the nuclei, and $\sigma_i$ is the total cross section per atom of the interaction. The detection efficiencies in both materials should be very close to 100\% at these low energies; thus, effects of a material and detector geometry dependent detection efficiency are neglected in the following. 

The total cross section will depend on the coupling of the dark matter particle to the detection media. For an electromagnetic conversion a Z$^2$ (where Z is the atomic number) scaling of the cross section is natural and is thus considered in the comparison of the rate limits in Ge from this analysis with the rate observed by DAMA. Another scaling can be trivially considered. This is a more general model than the one considered in our axion search paper \cite{cdmsaxion2008}. The scaled rate limits in NaI at a 90\% CL are given in \mbox{Fig.\,\ref{fig:upperlim}} (blue/dashed line). The total counting rate observed by DAMA/LIBRA is 6.8$\sigma$ greater than the upper limit at 3.15\,keV.

Under standard halo assumptions a conservative upper limit on the modulation amplitude is $\pm$6\% if the modulation is caused by a change in the particle flux only \cite{Lewin1995}. Note, that if the conversion cross section is inversely proportional to the dark matter particle velocity (as inelastic cross sections tend to be \cite{pospelov2008}) the annual modulation amplitude is highly suppressed. The insert in Fig.\,\ref{fig:upperlim} compares the unscaled upper limit (black/solid) and the Z$^2$ scaled upper limit in NaI (blue/dashed) on the modulation amplitude with the 2$\sigma$ regions of the annual modulation amplitude observed by DAMA (NaI+LIBRA) in the 2\,-\,4 keV (red/filled) and \mbox{2\,-\,6 keV} (green/hatched) energy range \cite{damalibra2008}. The upper limits on the modulation amplitudes are a factor of $\sim$2 less than observed by DAMA.


In this paper we reported on our analysis of the low-energy electron-recoil spectrum of the CDMS experiment, providing the observed rate in the \mbox{2\,-\,8.5 keV} range and the identification of possible background sources. The analysis found no significant excess in the counting rate above background. Considering the conversion of a dark matter particle into electromagnetic energy the 90\% CL upper limit on the total counting rate from CDMS at 3.15\,keV is 8.9$\sigma$ (6.8$\sigma$) below  the excess rate observed by DAMA in a direct comparison (under the assumption of a Z$^2$ scaling of the cross section), neglecting a possible background contribution from $^{40}$K in the DAMA data. We note that the actual scaling between Ge and NaI has to be provided by a specific model, but stress that an analysis of the low-energy electron-recoil spectrum of CDMS helps to identify or constrain possible models which can explain the annual modulation signature observed by DAMA. In the conservative case of a 6\% modulation amplitude our recent data provides 90\% CL upper limits on the modulation amplitude that are a factor of $\sim$2 less than observed by DAMA.

 This work is supported in part by the National Science Foundation (Grant Nos.\ AST-9978911, PHY-0542066, PHY-0503729, PHY-0503629,  PHY-0503641, PHY-0504224, PHY-0705052, PHY-0801536, PHY-0801708, PHY-0801712 and PHY-0802575), by the Department of Energy (Contracts DE-AC03-76SF00098, DE-FG02-91ER40688, DE-FG02-92ER40701, DE-FG03-90ER40569, and DE-FG03-91ER40618), by the Swiss National Foundation (SNF Grant No. 20-118119), and by NSERC Canada (Grant SAPIN 341314-07).


\begin{thebibliography}{99}

\bibitem{dama2000}
R.~Bernabei {\it et al.}, Phys.\ Lett. B {\bf 480}, 23 (2000).

\bibitem{damalibra2008}
  R.~Bernabei {\it et al.}, Eur.\,Phys.\,J.\,C {\bf56}, 333 (2008). 

\bibitem{cdms2008}
  Z.~Ahmed {\it et al.},  Phys.\ Rev.\ Lett.  {\bf 102}, 011301 (2009).

\bibitem{xenon10}J.~Angle {\it et al.},  Phys.\ Rev.\ Lett.  {\bf 100}, 021303 (2008).

\bibitem{xenon10sd}J.~Angle {\it et al.},  Phys.\ Rev.\ Lett.  {\bf 101}, 091301 (2008).


\bibitem{coupp}E.~Behnke {\it et al.}, Science {\bf 319}, 933 (2008).

\bibitem{kims}H.S.~Lee {\it et al.},  Phys.\ Rev.\ Lett.  {\bf 99}, 091301 (2007).

\bibitem{damaaxion}
R.~Bernabei {\it et al.}, Int.\,J. Mod.\, Phys. A {\bf21}, 1445 (2006). 

\bibitem{cogent}  C.~E.~Aalseth {\it et al.}, Phys.\ Rev.\ Lett  {\bf 101}, 251301 (2008).

\bibitem{cdmsaxion2008}
  Z.~Ahmed {\it et al.},
  arXiv:0902.4693v1 [hep-ex].


\bibitem{zips}
  K.~D.~Irwin {\it et al.}, Rev.\ Sci.\ Instr. {\bf 66}, 5322 (1995);
  T.~Saab {\it et al.}, AIP Proc. {\bf 605}, 497 (2002).

\bibitem{prd118} 
  D.~S.~Akerib {\it et al.}, Phys.\ Rev.\ D  {\bf 72}, 052009 (2005).
  
\bibitem{mei2009}D.-M.~Mei, Z.-B.~Yin and S.~R.~Elliot, arXiv:0903.2273v1 [nucl-ex].

\bibitem{james} 
F.~James, {\it Statistical Methods in Experimental Physics (2nd edition)}, World Scientific Publishing Co. Pte. Ltd. (2005).

\bibitem{damaapparatus}
R.~Bernabei {\it et al.}, Nucl.\,Instrum.\,Meth.\, A {\bf 592}, 297 (2008).
  
\bibitem{Lewin1995}
J.~D.~Lewin and P.~F.~Smith,  Astropart.\ Phys.\  {\bf 6}, 87 (1996).

\bibitem{pospelov2008}
  M.~Pospelov, A.~Ritz \& M.~Voloshin, Phys.\ Rev.\ D  {\bf 78}, 115012 (2008).

\bibitem{channel2007}
E.~M.~Drobyshevski, Phys.\ Lett. A {\bf 23} 3077 (2008).

\bibitem{damachannel2007}
R.~Bernabei {\it et al.}, Eur.\,Phys.\,J.\, C {\bf 53}, 205 (2008). 






\end{thebibliography}
\end{document}